\documentclass[12pt]{iopart}

\pdfoutput=1
\usepackage{iopams}
\usepackage{graphicx}
\usepackage{subfigure}
\begin{document}

\title{Mean-field analysis of the majority-vote model broken-ergodicity  steady state}

\author{ Paulo F.\ C.\ Tilles and Jos\'{e} F.\  Fontanari }

\address{Instituto de F\'{\i}sica de S{\~a}o Carlos,
            Universidade de S{\~a}o Paulo,
            Caixa Postal 369, 13560-970 S\~ao Carlos SP, Brazil}

\begin{abstract}
We study  analytically a variant of the  one-dimensional majority-vote model in which the individual retains its opinion in case there is a tie among the neighbors' opinions. The 
individuals are fixed in the sites of a ring of  size $L$ and can interact with their nearest neighbors only.
The interesting feature of this model is that it  exhibits an infinity of spatially heterogeneous absorbing configurations for $L \to \infty$ whose statistical properties we
probe analytically using a  mean-field framework  based on the decomposition of the $L$-site joint probability distribution into the $n$-contiguous-site joint distributions, the so-called 
$n$-site approximation.  To describe the  broken-ergodicity steady state  of the model we solve analytically   the mean-field dynamic equations for
arbitrary time $t$ in the cases $n=3$ and $4$. The asymptotic limit $t \to \infty$ reveals the mapping between the statistical properties of the random initial configurations and 
those of the final absorbing configurations. For the pair approximation ($n=2$) we derive that mapping using a trick that avoids solving the full dynamics.
Most remarkably, we find that the predictions of the $4$-site approximation reduce to those
of the $3$-site in the case of expectations involving three contiguous sites. In addition, those expectations fit the simulation data perfectly and so
we conjecture that they are in fact  the exact  expectations  for the one-dimensional majority-vote model.
  
\end{abstract}

\pacs{89.65.-s, 89.75.Fb, 87.23.Ge, 05.50.+q}

%
\section{Introduction} \label{sec:Intro}

A desirable property of a model for social behavior, or for complex systems in general,  is the presence of 
a nontrivial  steady state characterized by infinitely  many equilibrium points in the thermodynamic limit. This was the main
appeal of the  mean-field spin-glass models  used  widely since the 1980s to study  associative memory \cite{Hopfield_82},
prebiotic evolution \cite{Stein_84}, ecosystem organization \cite{deOliveira_02}, and social systems \cite{Lewenstein_92} to name just a few
of the areas impacted by the spin-glass approach to model complex systems \cite{Mezard_87}.

Models of social dynamics, however, are typically defined through the specification of the dynamic rules that govern the interactions between
agents \cite{RMP}  and so they are not amenable to analysis using tools borrowed from the equilibrium statistical mechanics of disordered  systems.
Nevertheless, the display  of a  steady state characterized by a multitude of  locally stable and  spatially inhomogeneous configurations remains a celebrated  feature of  this class 
of models, whose paradigm is Axelrod's model \cite{Axelrod_97},
since they can explain the diversity of cultures or opinions observed in human societies.  Axelrod's model is
attractive from the statistical physics perspective because it exhibits 
 a nonequilibrium phase transition  which separates  the spatially homogeneous (mono-cultural) from
the heterogeneous (multicultural) regimes \cite{Castellano_00,Vilone_02,Barbosa_09}. 

More recently, a long-familiar model of lattice statistical physics -- the majority-vote model \cite{Ligget_85,Gray_85} -- was  revisited in the context of social
dynamics models \cite{Parisi_03,Peres_10}. In fact, the majority-vote model  is a lattice version  of the classic frequency bias mechanism for opinion change
\cite{Boyd_85},  which assumes that the number of people holding an opinion is the key factor for an agent to adopt that opinion, i.e., people have a tendency 
to espouse opinions that are more common in their social environment. The variant of the majority-vote model considered in those studies  includes the state of
the target site (the voter) in the reckon of the majority (hence we refer to the model as extended majority-vote model), which happens to be the variant originally proposed in the physics 
literature \cite{Ligget_85,Gray_85}. This
fact is the sole responsible for the existence of an infinity of heterogeneous  absorbing configurations  whose statistical properties  were thoroughly studied via simulations 
in the case of a two-dimensional lattice \cite{Peres_10}.  Moreover, the non-linearity of the transition probabilities resulting from the inclusion of the
voter opinion in the majority reckoning makes the model not exactly solvable, in contrast to the voter model for which the transition probabilities are linear
\cite{Krapivsky_92}.

Many  interesting variants of the one-dimensional  majority-vote model were considered in the literature. For instance, some variants separate the individuals in  groups 
of fixed  sizes and apply the majority-vote rule to update  the opinion of the entire group  simultaneously \cite{Krapivsky_03}.  Others differentiate 
the groups a priori by introducing a group-specific  bias used to determine the  group opinion in case of ties \cite{Galam_05}.  Another variant of
interest is the non-conservative voter model for which the probability that the voter changes its opinion depends non-linearly on the fraction of
disagreeing neighbors \cite{Lambiotte_08}. In particular, this variant reduces to the model we study in this paper in the case the voter changes opinion solely when confronted by
a unanimity of opposite-opinion neighbors. All these variants of the majority-vote rule model were studied via simulations or within the single-site mean-field framework,
except for the non-conservative variant which was examined within the pair approximation as well \cite{Lambiotte_08}. Here we show that the single-site and
the pair approximations yield incorrect predictions for all statistical measures of the steady states and argue that the 3-site and 4-site approximations yield the exact
results for measures involving  up to three contiguous sites of the chain.

In this contribution we study analytically the one-dimensional version of the extended majority-vote model, which is described in Sect.\ \ref{sec:model}. Our goal was to understand how the
multiple-cluster steady state of  the model could  be described within the mean-field approach (Sect.\ \ref{sec:mean}). We find that the signature of the ergodicity breaking is
the appearance of an infinity of attractive fixed points  in the mean-field equations for the $n$-site approximation with $n\geq 2$.  The characterization
of the mean-field steady state requires the complete analytical solution of the dynamics in order to obtain  the mapping between the  statistical properties
of the random initial configurations and those  of the  final  absorbing configurations, except for the 2-site or pair-approximation for which we find a simple
shortcut to that mapping, as described in Sect.\ \ref{subsec:2}. The full solution of the dynamics was obtained  for the $3$ and $4$-site approximations in Sects.\ \ref{subsec:3} and
\ref{subsec:4}, respectively. We find that these two approximation schemes yield the very same expressions for the 
steady-state expectations involving three contiguous sites [see Eqs.\ (\ref{3.06}), (\ref{phi_3}) and (\ref{psi_3})]  and so we conjecture that those expressions are exact. 
A perfect fitting of the simulation data by those predictions adds further support to this claim. In addition we find that the steady-state expectations involving four contiguous sites
calculated within the 4-site approximation fit the simulation data perfectly. However, this approximation fails to describe higher order expectations.

\section{Model}\label{sec:model}

The agents are fixed at the sites of a ring of length $L$ and can interact with their nearest neighbors only.
The initial configuration is  chosen  randomly with  the opinion of each agent being specified by  
a random digit $1$ or $0$ with  probability $\rho_0$ and $1-\rho_0$, respectively. At each time we pick a target  agent at random and then verify
which is the more frequent opinion ($1$ or $0$) among its extended neighborhood, which includes the target agent itself.  The opinion of the target agent is then  changed 
to match the corresponding majority value. We note that there are no ties in the calculation of the preponderant opinion since the 
extended  neighborhood of any agent comprises exactly $3$ sites. As a result, the update rule of the model  is deterministic; stochasticity enters the model dynamics through
the choice of the target site and in the selection of the initial configuration.
The update procedure is repeated until 
the system is frozen in an absorbing configuration.

\begin{figure}[!t]
\centering\includegraphics{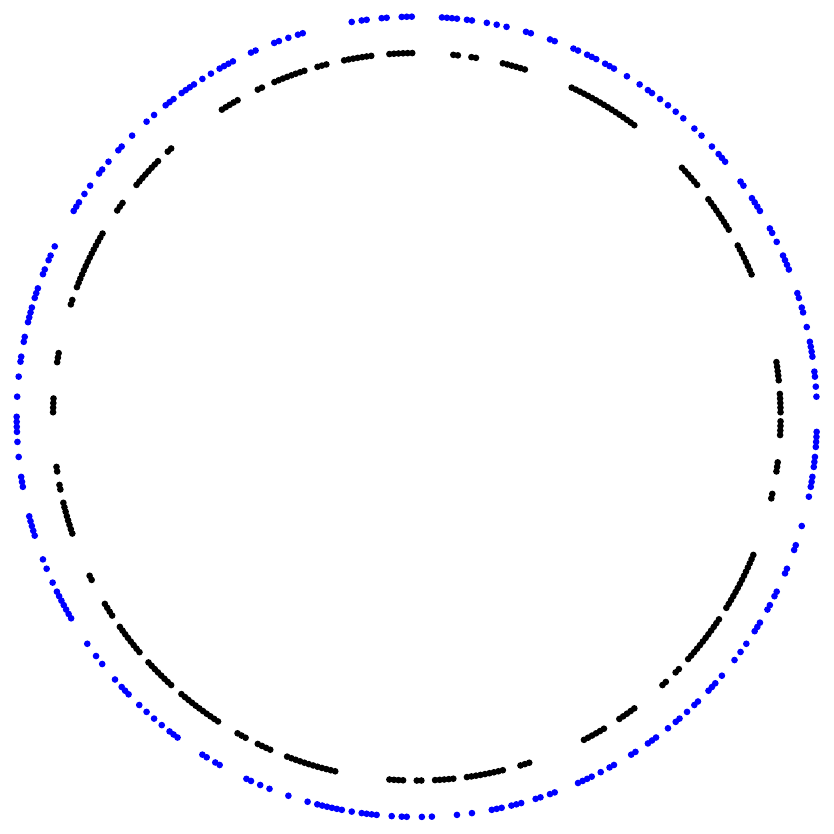}
\caption{(Color online) Disposition of the $\sigma_{i}=1$ variables in a ring with $L=500$ sites. The inner circle (black) is an absorbing configuration of the extended majority-vote model 
with $\rho =0.49$. There are 92 clusters and the largest cluster comprises 19 sites. The outer circle (blue) shows a random configuration with the same density of 1's. The total number of clusters is 255 
and the largest one comprises 9 sites.}
\label{fig:1}
\end{figure}

Although the majority-vote rule or, more generally, the frequency bias mechanism for
cultural change \cite{Boyd_85} is a homogenizing assumption by which the agents become more similar 
to each other, the two-dimensional version of the above-described model does  exhibit global polarization, i.e., a non-trivial stable 
multicultural regime  in the thermodynamic  limit \cite{Parisi_03,Galam_05,Peres_10}.  This regime should exist in the one-dimensional version
as well, since any sequence of two or more contiguous 1's (or 0's)  is stable under the update rule.
It should be noted that for 
the more popular variant of the majority-vote model, in which the state of the target site is not included in
the majority reckoning, and  ties are decided by choosing the opinion of the target agent at
random with probability $1/2$,  the only absorbing states in the thermodynamic limit are the two homogeneous configurations \cite{Tome_91,Oliveira_92}. 
As mentioned before,  the inclusion of the target site in the calculation of the majority is actually the original definition of  the
majority-vote model as introduced in Refs.  \cite{Ligget_85,Gray_85}. Figure \ref{fig:1} illustrates an absorbing configuration of the extended majority-vote model
together with a random configuration with the same density of 1's. The larger number of clusters (domains) observed in the random configuration is due to the
possibility of isolated sites, which are unstable under  the majority-vote rule.

As usual, we represent 
the state of the  agent at site  $i$ of the ring 
 by the binary variable $\sigma_i = 0,1$ and so the configuration of the entire ring  is 
denoted by $\sigma \equiv \left ( \sigma_1, \sigma_2, \ldots, \sigma_L \right )$. The master equation that governs the time evolution 
of the probability distribution $P \left ( \sigma, t \right )$ is given by
\begin{equation}\label{Master1}
\frac{d}{dt} P \left ( \sigma, t \right ) = \sum_i \left [ W_i \left ( \tilde{\sigma}^i \right ) P \left (\tilde{\sigma}^i, t \right )
- W_i \left ( \sigma \right ) P \left (\sigma, t \right )  \right ]
\end{equation}
where  $\tilde{\sigma}^i = \left ( \sigma_1, \ldots, 1- \sigma_i, \ldots, \sigma_L \right )$ and $W_i \left ( \sigma \right )$ is 
the transition rate between configurations $\sigma$ and $\tilde{\sigma}^i$ \cite{Tome_91,Oliveira_92}.
For the  one-dimensional extended majority-vote model  we have
\begin{equation}\label{W1}
W_{i} \left(\sigma \right) = \sigma_{i} \left(1 -\sigma_{i-1} -\sigma_{i+1}\right) + \sigma_{i-1}\sigma_{i+1}
\end{equation}
for $i=1, \ldots, L$. The boundary conditions
are such that $\sigma_0 = \sigma_L$ and  $\sigma_{L+1} = \sigma_1$. To implement  the $n$-site approximation up to $n=4$ we need to evaluate
the following expectations 
\begin{equation} \label{d1dt}
\frac{d}{dt}  \left \langle \sigma_i \right \rangle =   \left \langle \left ( 1 - 2 \sigma_i \right )  W_i \left ( \sigma \right )  \right \rangle ,
\end{equation}
\begin{equation} \label{d2dt}
\frac{d}{dt}  \left \langle \sigma_i \sigma_j  \right \rangle   =  2 \left \langle \sigma_j \left ( 1 - 2 \sigma_i \right )  W_i \left ( \sigma \right )    \right \rangle ,
\end{equation}
\begin{equation} \label{d3dt} 
\frac{d}{dt}  \left \langle \sigma_i \sigma_j \sigma_k  \right \rangle    =  3 \left \langle \sigma_j  \sigma_k \left ( 1 - 2 \sigma_i \right )  W_i \left ( \sigma \right )    \right \rangle ,
\end{equation}
\begin{equation} \label{d4dt} 
\frac{d}{dt}  \left \langle \sigma_i \sigma_j \sigma_k  \sigma_l \right \rangle    =  4 \left \langle \sigma_j  \sigma_k \sigma_l \left ( 1 - 2 \sigma_i \right )  W_i \left ( \sigma \right )    \right \rangle 
\end{equation}
where all indexes are assumed distinct and we have introduced the notation $ \left \langle  \left ( \ldots \right ) \right \rangle \equiv \sum_\sigma \left ( \ldots \right )  P \left (\sigma, t \right ) $.
The $n$-site approximation is  based on the calculation of this average by replacing  the full joint distribution probability  $P \left (\sigma, t \right )$ by  a decomposed   form that depends on the order $n$ of the 
approximation [see Eqs.\ (\ref{P1}), (\ref{P2}), (\ref{P3}) and (\ref{P4})].
Of course, in the derivation of Eqs.\ (\ref{d1dt})-(\ref{d4dt}), which generalize trivially to an arbitrary number of sites, we have assumed translational invariance, i.e., all sites are assumed equivalent.

\section{Mean-field  Analysis}\label{sec:mean}

In this section we  study the one-dimensional extended majority-vote model using the well-known  mean-field $n$-site approximation 
(see \cite{Konno_90, Avraham_92, Ferreira_93,Ferreira_09}). 
The basic idea behind the $n$-site approximation is to rewrite the  distribution $P \left ( \sigma, t \right )$ in terms of elementary joint probabilities
of $n$ contiguous sites only  and then deriving a system of self-consistent equations for these probabilities.  This key idea is expressed mathematically 
using the following equation which summarizes the  approximation scheme,
\begin{equation}\label{conditional}
P_{1\mid L-1} \left ( \sigma_i \mid \sigma_1, \ldots,  \sigma_L \right )  =  
P_{1\mid 2n-2} \left ( \sigma_i \mid \sigma_{i-n+1}, \ldots, \sigma_{i+n-1} \right )
\end{equation}
where, of course, $\sigma_i$ does not appear in the  arguments  at the right of the $\mid$ delimiter  in these conditionals.
This procedure  will be illustrated in the next subsections for $n=1$ to $n=4$. It is interesting to  note that, except for the single-site approximation $n=1$,
the states of any two sites are statistically dependent variables regardless of their positions in the ring.

\subsection{The single-site approximation}

This is the simplest mean-field scheme which assumes that the state of the agents at different  sites are independent random variables so that 
\begin{equation}\label{P1}
P \left ( \sigma, t \right ) = p_1\left ( \sigma_1, t \right ) p_1\left ( \sigma_2, t \right ) \ldots  p_1\left ( \sigma_L, t \right ) 
\end{equation}
and so  it is only necessary to calculate  the one-site distribution $ p_1\left ( \sigma_i, t \right )$ to describe the dynamics completely.
This can be done by noting that 
$ \rho \equiv \langle \sigma_i \rangle_t =  p_1\left ( 1, t \right )$ and using Eq.\ (\ref{d1dt}) to derive a self-consistent equation
for $\rho$. The final result is simply
\begin{equation}
\dot{\rho} = \rho \left(-2\rho^{2} + 3\rho -1\right) \label{1.02}
\end{equation}
with the notation $\dot{x} = dx/dt$. We note that $\rho$ contains the same information as the  single-site probability 
distribution $p_1 \left ( \sigma_i \right ) $ since $p_1 \left ( \sigma_i = 1 \right ) = \rho $ and  $p_1 \left ( \sigma_i = 0 \right ) = 1 -\rho $. 
A straightforward stability analysis shows that there are three fixed points,
\begin{equation}
\rho_{1} = 0, \quad \rho_{2} = 1/2, \quad \rho_{3} = 1,
\end{equation}
with only $\rho_{1}$ and $\rho_{3}$ being stable. This means that only the homogeneous configurations are stable and so the single-site approximation completely fails
to describe the steady state of the extended majority-vote model.
 
\subsection{The pair approximation}\label{subsec:2}

Using Eq.\ (\ref{conditional})  to write the full probability distribution in terms of the  joint probability of  two sites  and omitting the time dependence  we find 
\begin{equation}\label{P2}
P \left ( \sigma \right )  = \frac{p_2 \left ( \sigma_1, \sigma_2 \right ) p_2 \left ( \sigma_2, \sigma_3 \right )  \ldots p_2 \left ( \sigma_{L-1}, \sigma_L \right ) p_2 \left ( \sigma_{L}, \sigma_1 \right )}
{p_1 \left ( \sigma_1 \right ) p_1 \left ( \sigma_2 \right )\ldots  p_1 \left ( \sigma_{L-1} \right ) p_1 \left ( \sigma_L \right )}
\end{equation}
where $p_1  \left ( \sigma_i \right ) = \sum_{\sigma_j} p_2 \left ( \sigma_i, \sigma_j \right )$.  To avoid overburden the notation  we use the same notation for $p_1$ as  done in the 
single-site approximation  but the $p_1$ which appears in Eq.\ (\ref{P2}) is numerically distinct from that calculated in the previous section.  This simplifying convention for the notation 
of probabilities will be used in the next sections as well.

Within this framework, it is only necessary to calculate $p_2 \left ( \sigma_i, \sigma_{i+1} \right )$ to describe
the dynamics of the model completely.  This amounts to 4 variables, namely, $p_{2} \left(1,1\right), p_{2} \left(1,0\right), p_{2} \left(0,1\right)$ and $p_{2} \left(0,0\right)$, but
use of the normalization condition and of the parity symmetry [$p_{2} \left(1,0\right) = p_{2} \left(0,1\right)$] allows us to reduce the number of independent variables to  only 2.
The first  variable we pick is  $\phi  \equiv \left \langle \sigma_i \sigma_{i+1} \right \rangle =  p_{2} \left(1,1\right) $   which is given by Eq.\ (\ref{d2dt}) with $j=i+1$.  Next, noting that
$p_{2} \left(1,0\right) = p_{1} \left(1\right) - p_{2} \left(1,1\right) = \rho - \phi$ we pick $\rho$, given  by Eq.\ (\ref{d1dt}), as the second independent variable. Carrying out the
averages in the right-hand sides of Eqs.\ (\ref{d1dt}) and (\ref{d2dt})  using the decomposition (\ref{P2}) yields
\begin{equation}\label{r2}
 \dot{\rho} = \frac{\left( \rho -\phi \right) ^{2}\left( 2\rho -1\right)}{2\rho \left( 1-\rho \right)}
\end{equation}
and
\begin{equation}\label{f2}
\dot{\phi} = \frac{\left( \rho -\phi \right) ^{2}}{1-\rho } .
\end{equation}
The steady-state condition $\dot{\phi} = \dot{\rho} = 0$ as well as the numerical integration of these equations yields  $\rho = \phi$ for $t \to \infty$, with $\rho$ determined 
by the value of the initial condition $ \rho \left ( t=0 \right ) = \rho_0$ and $\phi \left ( t=0 \right ) = \rho_0^2$.   We note that this result implies that $p_{2}\left( 1,0\right) = p_{2}\left( 0,1\right) = 0$,
meaning that
the number of interfaces between clusters, i.e., of contiguous sites in different states at the steady state, is not extensive. This prediction is not correct as indicated by the higher-order
approximations and by the simulation data. Despite this incorrect  prediction, the pair approximation explains the most remarkable aspect of the extended majority-vote model, namely, 
the ergodicity breaking reflected by the infinity of
distinct absorbing  configurations.

\begin{figure}[!h]
\centering\includegraphics[width=0.6\linewidth]{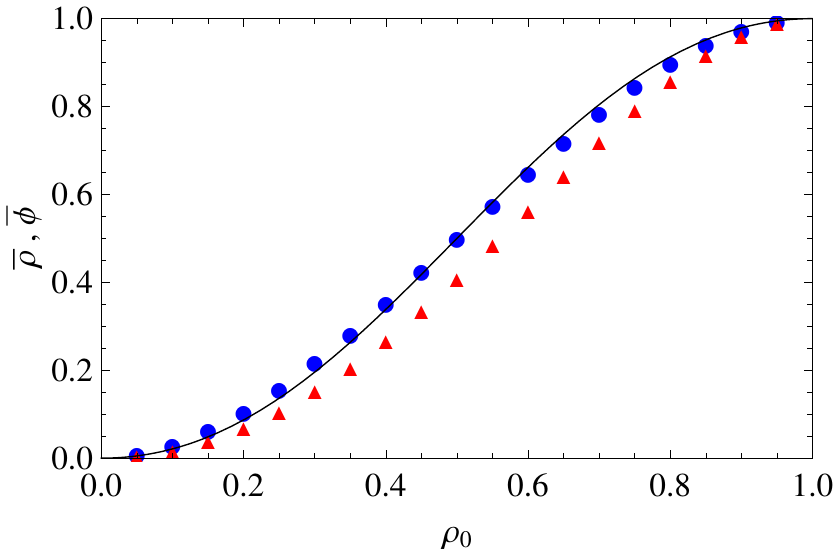}
\caption{(Color online) The fraction of sites in state $1$ at equilibrium, $\bar{\rho}$ (filled circles), and 
the probability that two neighbors are in state $1$ at equilibrium, $\bar{\phi}$ (filled triangles)
as functions of the initial fraction of sites in state $1$, $\rho_0$. 
The solid line is the result of the pair approximation for which $ \bar{\rho} = \bar{\phi} $  given  by Eq.\ (\ref{2.07}).
The initial condition is $\rho = \rho_0$ and $\phi = \rho_0^2$. The symbols
show the results of the  simulations for a ring of size $L=10^4$ and $10^6$ independent samples.}
\label{fig:2}
\end{figure}

The imposition of the steady-state condition is not sufficient to determine the equilibrium solution $\bar{\rho} = \bar{\phi}$ because there is a continuum of fixed points characterized by the function $\bar{\rho} \left(\rho_{0}\right)$.  In order to obtain this function or map, we revert to the original  variable $x \equiv p_2 \left ( 1, 0 \right ) =  \rho - \phi$, and rewrite Eqs.\ (\ref{r2}) and (\ref{f2}) as
\begin{equation}
\dot{\rho} = \frac{x^{2}\left( 2\rho -1\right)}{2\rho \left( 1-\rho \right)},
\end{equation}
\begin{equation}
\dot{x} = -  \frac{x^{2}}{2 \rho \left(1 - \rho\right) }.
\end{equation}
from which we can immediately obtain the integral equation
\begin{equation}
\int_{x_{0}}^{x\left(t\right)} d x^{\prime} = - \int_{\rho_{0}}^{\rho\left(t\right)} \frac{d\rho^{\prime}}{2\rho^{\prime}  - 1}. \label{2.06}
\end{equation}
As the stationary regime is obtained in the limit $t \rightarrow \infty$, we define $\bar{\rho} \left(\rho_{0}\right) \equiv \rho \left( t \rightarrow \infty\right)$. In addition, using $x_{0} = \rho_{0} \left(1 -\rho_{0} \right)$ and $x \left( t \rightarrow \infty\right) = 0$ (steady-state condition) we find
\begin{equation}
\bar{\rho} \left( \rho_{0}\right) = \frac{1}{2} \left[1 + \left(2 \rho_{0} -1\right) e^{2 \rho_{0} \left(1-\rho_{0} \right)} \right]. \label{2.07}
\end{equation}
This equation is identical to that derived for the  non-conservative voter model in the case the voter changes opinion only when confronted by
a unanimity of opposite-opinion neighbors\cite{Lambiotte_08}.
Figure\ \ref{fig:2} shows this steady-state solution together with the results of the simulations for a ring with $L=10^4$ sites. Despite the
incorrect prediction ($\bar{\phi} = \bar{\rho} $), Eq.\ (\ref{2.07}) yields a remarkably good quantitative agreement with the density $\rho$ obtained from
the simulations.  However, as we will show next the (supposedly) exact expression for $\bar{\rho} $ is much simpler than Eq.\ (\ref{2.07}).

\subsection{The $3$-site approximation}\label{subsec:3}

In this scheme the decomposition of $P \left ( \sigma \right )$ is 
\begin{equation}\label{P3}
P \left ( \sigma \right )  = \frac{p_3 \left ( \sigma_1, \sigma_2, \sigma_3\right ) p_3 \left ( \sigma_2, \sigma_3,\sigma_4 \right )  \ldots  p_3 \left ( \sigma_{L}, \sigma_{1},
\sigma_2 \right )}
{p_2 \left ( \sigma_1,\sigma_2 \right ) p_2 \left ( \sigma_2,\sigma_3 \right )\ldots  p_2 \left ( \sigma_L,\sigma_1 \right )}
\end{equation}
where $p_2  \left ( \sigma_i, \sigma_{i+1}\right ) = \sum_{\sigma_{i+2}} p_3 \left ( \sigma_i, \sigma_{i+1},\sigma_{i+2} \right )$. 
The goal here is to calculate the 9 probability values  $p_3 \left ( \sigma_i, \sigma_{i+1},\sigma_{i+2} \right )$ with $\sigma_k = 0,1$.
As before, use of the normalization condition  and of the parity symmetry  give us  6 variables to be 
determined using appropriate linear combinations of Eqs.\ (\ref{d1dt})-(\ref{d4dt}). We choose the following variables
\begin{equation}
\eqalign{x_{0} = p_{3}\left( 0,0,0\right) ,\qquad ~~x_{1}=p_{3}\left( 1,0,0\right), \qquad  ~x_{2} = p_{3}\left( 1,1,0\right), \cr
x_{1C}=p_{3}\left( 0,1,0\right), \qquad x_{2C}=p_{3}\left( 1,0,1\right), \qquad x_{3}=p_{3}\left(
1,1,1\right) .}  \label{3.01}
\end{equation}
which  are given by the expectations 
$x_1 = \left \langle \sigma_i \right \rangle - 
\left \langle \sigma_i \sigma_{i+1} \right \rangle - \left \langle \sigma_i \sigma_{i+2} \right \rangle + \left \langle \sigma_i \sigma_{i+1} \sigma_{i+2} \right \rangle$, 
 $x_2 = \left \langle \sigma_i \sigma_{i+1} \right \rangle - \left \langle \sigma_i \sigma_{i+1} \sigma_{i+2} \right \rangle $,
  $x_3 = \left \langle \sigma_i \sigma_{i+1} \sigma_{i+2} \right \rangle $,
and so on.

\subsubsection{Mean-field equations.}  Evaluating the averages in Eqs.\ (\ref{d1dt})-(\ref{d4dt})
using the decomposition  (\ref{P3}) yields
\begin{equation}
\eqalign{\hspace{-1.5cm} \dot{x}_{1} = \frac{1}{3} \frac{x_{1C}}{x_{1C}+x_{2}} \left( x_{2C} - x_{1} \right) , \qquad ~~~~~~~~~
\dot{x}_{2} = \frac{1}{3} \frac{x_{2C}}{x_{2C}+x_{1}} \left(x_{1C} - x_{2}\right), \cr
\hspace{-1.5cm} \dot{x}_{1C} = -\frac{1}{3} \frac{x_{1C}}{x_{2C}+x_{1}} \left(3 x_{2C} + x_{1} \right) , \qquad ~~~ \dot{x}_{2C} = -\frac{1}{3} \frac{x_{2C}}{x_{1C}+x_{2}} \left(3 x_{1C} + x_{2}\right), \cr
\hspace{-1.5cm} \dot{x}_{0} = \frac{1}{3} \frac{x_{1C}}{x_{1C}+x_{2}} \left(x_{1C} + 2 x_{1} + x_{2} \right), \qquad  \dot{x}_{3} = \frac{1}{3} \frac{x_{2C}}{x_{2C}+x_{1}} \left(x_{2C} + x_{1} + 2 x_{2} \right).
}\label{3.02}
\end{equation}
The steady state  is given by  $x_{1C} = x_{2C} = 0$,  i.e., $p_3 \left ( 0,1,0 \right ) =p_3 \left ( 1,0,1 \right ) = 0$ which, in contrast to the situation we found in the analysis of
the pair approximation,  reflects the physical requirement that absorbing configurations cannot exhibit  isolated sites. As we are still left with $4$ undetermined  variables after applying 
the steady-state condition, we need an alternative
method to characterize the steady state. Somewhat surprisingly, in this case we will be able to solve the dynamics analytically, a feat that seems unfeasible in the case of the pair approximation.  

In fact, what makes the system of nonlinear coupled equations (\ref{3.02}) solvable is the observation that $ y \equiv  x_{1C} + x_{2} = x_{2C} + x_{1} = p_2 \left ( 1,0 \right ) $, so
the denominators in the r.h.s. of all those equations are identical. In addition,  we note that $x_0$ does not affect the other 5 variables so we can first solve for them and then return to the
equation for $x_0$ to complete the solution of the  system (\ref{3.02}).

\begin{figure}[!h]
\centering\includegraphics[width=0.6\linewidth]{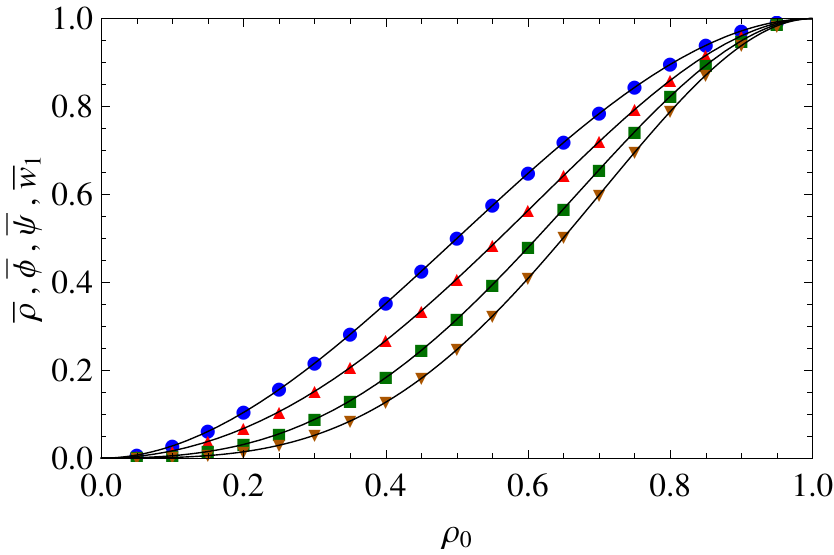}
\caption{(Color online)  The solid lines are the analytical results for the steady-state measures obtained with the $4$-site approximation while the symbols represent the 
results of the  simulations for a ring of size $L=10^4$ and $10^6$ independent samples. The convention is (from top to bottom) $\bar{\rho} \left(\rho_{0} \right)$ (circles), $\bar{\phi} \left(\rho_{0} \right)$ (triangles), 
$\bar{\psi} \left(\rho_{0} \right)$ (squares) and $\bar{w}_{1} \left(\rho_{0} \right)$ (upside down triangles). The upper three curves are identical for the $3$-site approximation.}
\label{fig:3}
\end{figure}

Introducing the auxiliary  variables $z_{1} = x_{1C} + x_{2C}$, $z_{2} = x_{1C} - x_{2C}$, and recalling that  $\rho = x_{1} + x_{2} + x_{2C} + x_{3}$ we reduce  Eqs.\ (\ref{3.02})
to 
\begin{equation} \label{3.04}
\dot{y} = - \frac{z_{1}}{3}, \qquad \dot{z}_{1} = - \frac{z_{1}}{3} - \frac{z_{1}^{2} -z_{2}^{2}}{3 y}, \qquad \dot{z}_{2} = - \frac{z_{2}}{3}, \qquad \dot{\rho} = - \frac{z_{2}}{3},
\end{equation}
where we have omitted the equation for $x_0$.
The last two equations can be immediately solved  and yield
\begin{equation}\label{z2_3s}
z_{2} \left( t\right) = \rho_{0}\left(1 -\rho_{0}\right)\left(1 -2\rho_{0}\right) e^{-\frac{1}{3}t} , 
\end{equation}
\begin{equation}
\rho \left( t\right) =   \rho_{0}^{2} \left(3 - 2\rho_{0} \right) + \rho_{0}\left(1 -\rho_{0}\right)\left(1 -2\rho_{0}\right)  e^{-\frac{1}{3}t} , \label{3.05}
\end{equation}
 Hence in the asymptotic limit $t \rightarrow \infty$ we obtain 
\begin{equation}\label{3.06}
\bar{\rho} \left( \rho_{0}\right) = \rho_{0}^{2} \left(3 - 2\rho_{0} \right)  
\end{equation}
and $\bar{z}_2 = 0$, as expected, since  both $x_{1C} $ and $x_{2C}$ vanish at the steady state. 
For $\rho_0 \to 0$  or $\rho_0 \to 1$ the pair approximation 
estimate of $\bar{\rho}$ given by  Eq.\ (\ref {2.07})
reduces to Eq.\  (\ref{3.06}) in a first order approximation in $\rho_0$ or $1 - \rho_0$.  Equation   (\ref{3.06})   describes the simulation data perfectly as illustrated in Fig. \ref{fig:3} and,
as already mentioned, we believe it gives the exact  value for the steady-state density of 1s of the one-dimensional majority-vote model.

The  explicit calculation of the remaining two unknowns $y$ and $z_1$ using Eqs.\  (\ref{3.04}) is not  too involved and their knowledge will allow us to
evaluate  other quantities of interest,  such as $\phi$ and other high-order correlations.
We begin by introducing the auxiliary variables $\omega_{1} = z_{1}/y$ and $\omega_{2} = z_{2}/y$ which satisfy the equations
\begin{equation}
\dot{\omega}_{1} = -\frac{1}{3} \omega_{1} + \frac{1}{3}  \omega_{2}^{2}, \qquad \dot{\omega}_{2} = -\frac{1}{3} \omega_{2} + \frac{1}{3} \omega_{1} \omega_{2}. \label{3.07}
\end{equation}
Next,  we define $\alpha = \omega_{1}^{2} - \omega_{2}^{2}$  which is given by  $\dot{\alpha} = -2 \alpha/3$ and so
\begin{equation}
\alpha \left(t\right) = \alpha_{0} e^{-\frac{2t}{3}}, \label{3.08}
\end{equation}
with $\alpha_{0} = 4 \rho_{0} \left(1 -\rho_{0}\right)$. At this point we can readily write an explicit equation for $z_{1}$ in terms of $y$,
\begin{equation}
z_{1} \left(t\right)= e^{-\frac{t}{3}}  \sqrt{z_{0}^{2} +\alpha_{0}  y^{2}\left(t\right)}, \label{3.09}
\end{equation}
where we used the fact that  $z_{1} \geq 0$ and  that $z_{2}$ is given by Eq.\ (\ref{z2_3s}). Now,   inserting this expression in the equation for $y$ [see Eq.\  (\ref{3.04})]  results in an easily solvable integral,
\begin{equation}
\int_{y_{0}}^{y} \frac{d y^{\prime}}{\sqrt{m_{0}^{2} + y^{\prime 2}}} = \log{\left(\frac{y + \sqrt{m_{0}^{2}+y^{2}}}{y_{0} + \sqrt{m_{0}^{2}+y_{0}^{2}}} \right) }, \label{3.10}
\end{equation}
with $m_0^2 = z_0^2/\alpha_0$ and $y_{0} = \rho_{0}\left( 1-\rho_{0}\right)$.  Finally, carrying out the integration yields 
\begin{eqnarray}
y \left( t\right) = &\frac{1}{2} \left(y_{0} + \sqrt{m_{0}^{2}+y_{0}^{2}} \right) \exp{\left[ -\sqrt{\alpha_{0}} \left(1 - e^{-\frac{t}{3}} \right) \right]} \nonumber \\
&- \frac{1}{2} \frac{ m_{0}^{2}}{y_{0} + \sqrt{m_{0}^{2}+y_{0}^{2}}} \exp{\left[ \sqrt{\alpha_{0}} \left(1 -e^{-\frac{t}{3}}\right) \right]}. \label{3.11}
\end{eqnarray}
In the limit $t \to \infty$ this equation reduces to 
\begin{eqnarray}\label{3.18}
 \bar{y} \left(\rho_{0}\right) & = & \rho_{0} \left(1 -\rho_{0}\right) \cosh{\left[2 \sqrt{\rho_{0} \left(1 -\rho_{0}\right)}\right]} \nonumber \\
&  &    - \frac{1}{2} \sqrt{\rho_{0} \left(1 -\rho_{0}\right)} \sinh{\left[2 \sqrt{\rho_{0} \left(1 -\rho_{0}\right)}\right]} 
\end{eqnarray}
which exhibits the symmetry  $ \bar{y} \left(\rho_{0}\right) =  \bar{y} \left(1 - \rho_{0}\right) $.

To conclude the solution of the  system of equations  (\ref{3.02}) we need now to  determine $x_0$. The easiest way to do that
is to rewrite the equation for $x_0$ in (\ref{3.02}) as 
\begin{equation}
\dot{x}_{0} = -2\dot{y} +\frac{1}{2} \left(\dot{z}_{1} -3\dot{z}_{2} \right)  \label{3.14}
\end{equation}
which can be immediately integrated to  yield
\begin{equation}\label{3.15}
x_0 \left ( t \right ) = x_{0}\left(0\right) - 2 \left [y\left(t\right)  - y_{0} \right ] +\frac{1}{2} \left[z_{1}\left(t\right)  - z_{1}\left(0\right) \right ] 
-\frac{3}{2} \left[z_{2}\left(t\right)  - z_{2}\left(0\right) \right ] .
\end{equation}
In the asymptotic limit  $ x_0 \left ( t \to \infty \right) \equiv  \bar{\psi}_{-1} $ we find
\begin{equation}\label{psi-1}
\bar{\psi}_{-1} \left(\rho_{0}\right) = \left(1 +2\rho_{0}\right) \left(1 -\rho_{0}\right)^{2} -2\bar{y} \left(\rho_{0}\right) .
\end{equation}

\subsubsection{Simple steady-sate expectations.} At this stage we should be able to express any  quantity characterizing the ring configuration  at time $t$ 
in  terms of  the time-dependent variables $\rho$, $z_2$, $y$ and $z_1$. However, we will focus here only in the steady-state regime ($t \to \infty$) for which only 
$\rho$ and $y$ contribute since $\bar{z}_1 = \bar{z}_2 = 0$.

The  most interesting expectations are those whose time evolution are  defined by Eqs.\  (\ref{d1dt})-(\ref{d4dt}) in the case the indexes are associated to contiguous 
sites. We begin with 
 $\phi = \left \langle \sigma_i \sigma_{i+1} \right \rangle = x_2 +x_3$, and recall that $\rho = \left \langle \sigma_i  \right \rangle  = x_1 + x_2 + x_{2C} + x_3$ 
and  $y = x_1 + x_{2C}$ so that  $\phi = \rho - y$. Then at the steady state we find
\begin{equation}\label{phi_3}
\bar{\phi} \left(\rho_{0}\right)= \rho_{0}^{2} \left(3 - 2\rho_{0} \right) - \bar{y} \left(\rho_{0}\right) .
\end{equation}
Next we note that  $\psi \equiv \left \langle \sigma_i \sigma_{i+1} \sigma_{i+2} \right \rangle  = x_3 = \rho - 2y +
\left ( z_1 + z_2 \right ) /2$ and so
\begin{equation}\label{psi_3}
\bar{\psi} \left(\rho_{0}\right) = \rho_{0}^{2} \left(3 - 2\rho_{0} \right) - 2\bar{y} \left(\rho_{0}\right) .
\end{equation}
These two steady-state expectations, which are shown in Fig.\ \ref{fig:3}, describe the simulation data perfectly. Expectations
involving more than three contiguous sites  must be decomposed so as to be described by the $3$-site approximation. Of particular interest is
the 4-site expectation $w_1  \equiv  \left \langle \sigma_i \sigma_{i+1} \sigma_{i+2} \sigma_{i+3} \right \rangle = p_{4} \left( 1,1,1,1\right)$,
which in the 3-site approximation scheme becomes  $w_1 = p_3^2 \left( 1,1,1\right) / p_2 \left ( 1, 1\right ) $ so that 
$\bar{w}_1 \left(\rho_{0}\right) = \bar{\psi}^{2}/\bar{\phi}$ at the steady state. In the scale of fig.\ \ref{fig:3} this result is indistinguishable from
the simulation data or from their counterpart calculated with the  $4$-site approximation (see fig.\ \ref{fig:4}).

For completeness, let us calculate $\phi_{-1} \equiv p_2 \left ( 0,0 \right ) = x_1 + x_0$ at the steady state. Since $x_0$ is given by Eq.\ (\ref{psi-1}) and
$x_1 =    y  - \left(z_{1} -z_{2}\right)/2$ we find 
\begin{equation}\label{phi-1}
\bar{\phi}_{-1} \left(\rho_{0}\right) = \left(1 +2\rho_{0}\right) \left(1 -\rho_{0}\right)^{2} -\bar{y} \left(\rho_{0}\right) .
\end{equation}
The fact that the dynamics is invariant to the interchange of 1s and 0s provided that we change $\rho_0 $ to $ 1- \rho_0$ is 
expressed by the easily verifiable identities $ \bar{\rho} \left( 1 - \rho_{0}\right) =1 - \bar{\rho} \left( \rho_{0}\right)$,
$\bar{\phi} \left(\rho_{0}\right) = \bar{\phi}_{-1} \left(1 -\rho_{0}\right)$, and  $\bar{\psi} \left(\rho_{0}\right) = \bar{\psi}_{-1} \left(1 -\rho_{0}\right) $.
Of course, this symmetry holds for all orders $n$ of the $n$-site approximation scheme and we will resort to it  to abbreviate the calculations of
the $4$-site approximation in Sect.\ \ref{subsec:4}.

\subsubsection{Probability of clusters of length $m$.}

A more informative quantity is the  probability of finding  a cluster of  $m > 1$ sites in an absorbing configuration.  There are only two possibilities
for such a cluster: ({\it a})  a site in state $\sigma_i = 0$  followed by $m$ sites in states $\sigma_{i+1} = \sigma_{i+2}  \ldots  \sigma_{i+m} = 1$ which are then
followed by another site in state $\sigma_{i+m+1} = 0$ and  ({\it b})  a site in state $\sigma_i = 1$  followed by $m$ sites in states $\sigma_{i+1} = \sigma_{i+2}  \ldots  \sigma_{i+m} = 0$ which are then
followed by another site in state $\sigma_{i+m+1} = 1$. The probability of these configurations happening in an absorbing configuration  can be easily derived using
the decomposition (\ref{P3}) and yields
\begin{equation}\label{P3cl1}
P^{(3)}_{cl} \left ( m \right ) =  \frac{ p_3^2 \left ( 0,1,1 \right ) }{p_2 \left ( 1,1 \right )}
\left [ \frac{p_3 \left ( 1,1,1 \right )}{p_2 \left ( 1,1 \right )} \right ]^{m-2} + 
\frac{p_3^2 \left ( 1,0,0 \right ) }{p_2 \left ( 0,0 \right )}\left [ \frac{p_3 \left ( 0,0,0 \right )}{p_2 \left ( 0,0 \right )} \right ]^{m-2}.
\end{equation}
To rewrite this expression in terms of more elementary  steady-state quantities we recall 
that $p_3 \left ( 1,0,0 \right ) = p_2 \left ( 0,0 \right ) -  p_3 \left ( 0,0,0 \right ) $ and  $p_3 \left ( 0,1,1 \right ) =  p_2 \left ( 1,1 \right ) - p_3 \left ( 1,1,1 \right ) $
so that
\begin{equation}\label{P3cl2}
P^{(3)}_{cl} \left (\rho_0,  m \right ) =
 \frac{\left( \bar{\phi} - \bar{\psi}\right)^{2}}{\bar{\phi}} \left( \frac{\bar{\psi}}{\bar{\phi}} \right)^{m-2} + \frac{\left( \bar{\phi}_{-1} -\bar{\psi}_{-1}\right)^{2}}{\bar{\phi}_{-1}} \left( \frac{\bar{\psi}_{-1}}{\bar{\phi}_{-1}} \right)^{m-2} .
\end{equation}
The 3-site approximation estimate for the  probability of finding clusters of length $m$ given by this equation  is presented in Figs.\ \ref{fig:5} and \ref{fig:6} together
with the results of the  simulations and the estimate of the 4-site approximation. We will postpone the  discussion of the physical implications of  the results
presented in these figures to Sect.\  \ref{sec:Discussion}.

\subsubsection{Two-site correlations.} Knowledge  of the two-site correlations defined by
\begin{equation}\label{def_corr}
\textrm{corr} \left(\sigma_{i},\sigma_{i+j}\right) = \langle \sigma_{i} \sigma_{i+j}\rangle - \langle \sigma_{i}\rangle \langle \sigma_{i+j}\rangle
\end{equation}
is very  useful to determine the validity of the approximations. Since all sites are equivalent we have $\langle \sigma_{i}\rangle = \langle \sigma_{i+j}\rangle = \bar{\rho} \left(\rho_{0}\right)$
regardless of the order $n$ of the approximation. Some two-site expectations follow straightforwardly from the previous results, namely, 
$\langle \sigma_{i} \sigma_{i}\rangle_{\left(3\right)} =  \bar{\rho}$, $ \langle \sigma_{i} \sigma_{i+1}\rangle_{\left(3\right)} =  \bar{\phi}$ and 
$\langle \sigma_{i} \sigma_{i+2}\rangle_{\left(3\right)} =  \bar{\psi}$. The first nontrivial two-site expectation is 
\begin{eqnarray}\label{corr_3_3}
\left \langle \sigma_{i} \sigma_{i+3} \right \rangle_{\left(3\right)} & = &  \sum_{\sigma_{i+1},\sigma_{i+2}} P \left ( \sigma_{i}=1,\sigma_{i+1},\sigma_{i+2},\sigma_{i+3}=1 \right )
\nonumber \\
& = &  \frac{\bar{\psi}^{2}}{\bar{\phi}} + \frac{\left(\bar{\phi}_{-1} -\bar{\psi}_{-1} \right)^{2}}{\bar{\phi}_{-1}}
\end{eqnarray}
where we have decomposed the 4-site probability  in terms of the elementary 3-site probabilities. Note that the sum has  two non-vanishing terms only, since
$P \left ( 1,0,1,1 \right ) = P \left ( 1,1,0,1 \right ) = 0$. Applying the very  same procedure  to  calculate $ \left \langle \sigma_{i} \sigma_{i+4} \right \rangle $ yields
\begin{equation}\label{corr_3_4}
\left \langle \sigma_{i} \sigma_{i+4} \right \rangle_{\left(3\right)} = 
\left(\frac{\bar{\psi}}{\bar{\phi}}\right)^{2} \bar{\psi} + 2  \frac{\left(\bar{\phi}_{-1} -\bar{\psi}_{-1} \right)^{2}}{\bar{\phi}_{-1}} + 
\left(\frac{\bar{\phi}_{-1} -\bar{\psi}_{-1}}{\bar{\phi}_{-1} }\right)^{2} \bar{\psi}_{-1} .
\end{equation}
In this case only 4  terms give nonzero contributions to the sum over the  middle sites. Equations (\ref{corr_3_3}) and (\ref{corr_3_4})  clarify a
fact that is often unappreciated, namely,  regardless of their position in the ring, the sites are always treated as statistically dependent variables within
the $n$-site approximation scheme for $n>1$.

\subsection{The $4$-site approximation}\label{subsec:4}

In the $4$-site approximation framework the decomposition of $P \left ( \sigma \right )$  is given by the prescription
\begin{equation}\label{P4}
P \left ( \sigma \right )  = \frac{p_4 \left ( \sigma_1, \sigma_2, \sigma_3, \sigma_4\right ) p_4 \left ( \sigma_2, \sigma_3,\sigma_4, \sigma_5 \right )  
\ldots  p_4 \left ( \sigma_{L},\sigma_1, \sigma_2, \sigma_3  \right )}
{p_3 \left ( \sigma_1,\sigma_2,\sigma_3 \right ) p_3 \left ( \sigma_2,\sigma_3, \sigma_4\right )\ldots  p_3 \left ( \sigma_L,\sigma_1, \sigma_2 \right )}
\end{equation}
where $p_3 \left ( \sigma_{i},\sigma_{i+1},\sigma_{i+2} \right ) = \sum_{\sigma_{i+3}} p_4 \left ( \sigma_{i},\sigma_{i+1},\sigma_{i+2}, \sigma_{i+3} \right )$.
Full determination of the joint probability  $ p_4 \left ( \sigma_{i},\sigma_{i+1},\sigma_{i+2}, \sigma_{i+3} \right )$ requires the calculation of 16 unknowns, but the normalization condition
and the parity symmetry allows us to reduce this number to the following 10  unknowns
\begin{eqnarray}\label{4.01}
\eqalign{\fl  w_{-1} = p_{4}\left( 0,0,0,0\right), \\
\fl x_{1} = p_{4}\left( 1,0,0,0\right) ,\quad x_{2}=p_{4}\left( 0,1,0,0\right), \\
\fl y_{1} = p_{4}\left( 1,1,0,0\right) ,\quad y_{2}=p_{4}\left( 1,0,1,0\right), \quad y_{3} = p_{4}\left( 1,0,0,1\right), \quad y_{4}=p_{4}\left( 0,1,1,0\right), \\
\fl z_{1} = p_{4}\left( 1,1,1,0\right), \quad z_{2}=p_{4}\left(1,1,0,1\right), \\
\fl  w_{1} =p_{4}\left( 1,1,1,1\right).} 
\end{eqnarray}
As usual,  Eqs. (\ref{d1dt})-(\ref{d4dt}) allow us to derive the equations for all these unknowns. For example, $z_1 = \langle \sigma_i \sigma_{i+1} \sigma_{i+2}\rangle
- \langle \sigma_i \sigma_{i+1} \sigma_{i+2} \sigma_{i+3}\rangle$, $w_1 = \langle \sigma_i \sigma_{i+1} \sigma_{i+2} \sigma_{i+3}\rangle$ and so on.

\subsubsection{Mean-field equation} Evaluation of the averages in  Eqs. (\ref{d1dt})-(\ref{d4dt})  using the decomposition (\ref{P4}) results in the following set of equations

\begin{eqnarray}\label{4.02}
\eqalign{\hspace{-1.0cm} \dot{w}_{-1} = \frac{x_{2}}{2} \left(1 + \frac{x_{1}}{x_{2} + y_{1}} \right), \\
\hspace{-1.0cm} \dot{x}_{1} = \frac{1}{4} \left[ \frac{x_{2}}{x_{2}+y_{1}} \left(y_{3} -x_{1} \right) + y_{2}\right], \qquad \dot{x}_{2} = \frac{1}{4} \left( \frac{y_{2}^{2}}{y_{2}+z_{2}} - \frac{x_{2} y_{2}}{x_{2}+y_{2}} - x_{2}\right),  \\
\hspace{-1.0cm} \dot{y}_{1} = \frac{y_{2}}{4} \left( \frac{x_{2} }{x_{2}+y_{2}} + \frac{z_{2} }{y_{2}+z_{2}}\right), 
\qquad  ~~~ \dot{y}_{2} = - \frac{y_{2}}{4} \left(   \frac{y_{2} }{x_{2}+y_{2}} + \frac{y_{2} }{y_{2}+z_{2}} + 2\right), \\
\hspace{-1.0cm} \dot{y}_{3} = - \frac{1}{2} \frac{x_{2} y_{3}}{x_{2} + y_{1}}, 
\qquad  ~~~~~~~ ~~~~~~~~~~~~~~\dot{y}_{4} = - \frac{1}{2} \frac{y_{4} z_{2}}{y_{1} + z_{2}},  \\
\hspace{-1.0cm} \dot{z}_{1} = \frac{1}{4} \left[ \frac{z_{2}}{y_{1} + z_{2}} \left( y_{4} - z_{1}\right) + y_{2} \right], 
\qquad  ~ \dot{z}_{2} = \frac{1}{4} \left( \frac{y_{2}^{2}}{x_{2} + y_{2}} - \frac{y_{2}z_{2}}{y_{2} + z_{2}} - z_{2}\right),  \\
\hspace{-1.0cm}  \dot{w}_{1} = \frac{z_{2}}{2} \left(1 + \frac{z_{1}}{y_{1} + z_{2}} \right).} 
\end{eqnarray}
As the variables $w_{-1}$ and $w_1$ do not affect the other variables and  the consistency conditions
\numparts 
\begin{eqnarray}
 \sum_{\sigma_{x_{1}},\sigma_{x_{2}}} p_{4} \left(1,1,\sigma_{x_{1}},\sigma_{x_{2}} \right)  & = & 
\sum_{\sigma_{x_{1}},\sigma_{x_{2}}} p_{4} \left( \sigma_{x_{1}},1,1,\sigma_{x_{2}} \right), \label{4.03a} \\ 
\sum_{\sigma_{x_{1}},\sigma_{x_{2}}} p_{4} \left(1,0,\sigma_{x_{1}},\sigma_{x_{2}} \right) & = & 
\sum_{\sigma_{x_{1}},\sigma_{x_{2}}} p_{4} \left(\sigma_{x_{1}},1,0,\sigma_{x_{2}} \right), \label{4.03b}
\end{eqnarray} 
\endnumparts
result in the simple relations,
\begin{equation}
y_{1} + z_{2} = y_{4} + z_{1}, \qquad x_{1} + y_{3} = x_{2} + y_{1}, \label{4.04}
\end{equation}
from where we can eliminate $x_{2}$ and $y_{1}$, we are actually left with a system of 6 coupled equations for
the variables $x_1, y_1,y_3,y_4,z_1$ and $z_2$.
(We note that Eqs.\ (\ref{4.03a})  and (\ref{4.03b}) merely exhibit  alternative ways of expressing $p_2 \left ( 1,1 \right )$ and  $p_2 \left ( 1,0 \right )$,
respectively.)

Introducing the linear transformation
\begin{equation}
\alpha_{\pm} = y_{4} \pm z_{1}, \qquad \eta_{\pm} = y_{2} \pm z_{2}, \qquad \delta_{\pm} = y_{3} \pm x_{1}, \label{4.05}
\end{equation}
we obtain the closed set of equations
\begin{eqnarray}\label{4.06}
\eqalign{\fl \dot{\alpha}_{+} = \frac{1}{4} \eta_{-}, 
\qquad ~~~~~~~~~~~~~~~~ \dot{\alpha}_{-} = - \frac{1}{4} \left[\eta_{+} + \frac{\alpha_{-}}{\alpha_{+}} \left(
\eta_{+} - \eta_{-}\right) \right], \\
\fl \dot{\eta}_{+} =  - \frac{1}{4} \left( 2\eta_{+} + \eta_{-}\right), 
\qquad  ~~~ \dot{\eta}_{-} =  - \frac{1}{8} \left[ \frac{\eta_{+}^{2}
+4\eta_{+} \eta_{-} +\eta_{-}^{2}}{\eta_{+}} - \frac{\left(\eta_{+} + \eta_{-} \right)^{2}}{\alpha_{+} - \eta_{+} - \delta_{+}}\right], \\
\fl   \dot{\delta}_{+} = \frac{1}{4} \left( \alpha_{+} + \eta_{-} - \delta_{+}\right), \qquad \dot{\delta}_{-} = \frac{1}{4} \left[ \alpha_{+}
-\eta_{+} -\delta_{+} + \frac{\delta_{-}}{\delta_{+}} \left( 2 \alpha_{+} - 2\delta_{+} -\eta_{+} +\eta_{-}\right)\right].} 
\end{eqnarray}
Although this system might look formidable, its solution is not very involved.  We begin by eliminating $\eta_-$ in the equations for
$\alpha_{+}$ and $\delta_{+}$ in order to get
\begin{equation}
\dot{\delta}_{+} + \frac{1}{4} \delta_{+} = \dot{\alpha}_{+} + \frac{1}{4} \alpha_{+}. \label{4.07}
\end{equation}
The auxiliary variable $f = \alpha_{+} -\delta_{+}$ satisfies $\dot{f} = -\frac{1}{4}f$, whose solution is
\begin{equation}\label{4.08}
f \left( t\right) = \left(\alpha_{+0} -\delta_{+0}\right) e^{-\frac{t}{4}} = \rho_{0} \left( 1 -\rho_{0}\right) \left( 2\rho_{0} -1\right) e^{-\frac{t}{4}}. 
\end{equation}
This explicit solution for $f$ in terms of $\rho_0$ and $t$ allows us to consider the equations for $\eta_{\pm}$ as a closed subset of equations which
can be solved as follows. The change of variables 
\begin{equation}
\omega = \frac{\eta_{+} +\eta_{-}}{\eta_{+}}, \qquad \gamma = \frac{\eta_{+} +\eta_{-}}{f -\eta_{+}},  \label{4.09}
\end{equation}
leads to the much simpler equations
\begin{equation}
\dot{\omega} = -\frac{1}{4} \omega +\frac{1}{8} \omega^{2} +\frac{1}{8} \gamma \omega , \qquad
\dot{\gamma} = -\frac{1}{4} \gamma -\frac{1}{8} \gamma^{2} -\frac{1}{8} \gamma \omega  \label{4.10}
\end{equation}
which imply that $ d \left( \gamma \omega\right)/dt = - \frac{1}{2} \gamma \omega$. Hence
\begin{equation}
\gamma \left( t\right) \omega\left( t\right) = -4 \rho_{0} \left(1 -\rho_{0}\right) e^{-\frac{t}{2}},  \label{4.11}
\end{equation}
where we have used $\gamma \left( t=0\right) = -2\rho_{0}$ and $\omega\left( t = 0 \right) = 2\left(1 -\rho_{0}\right)$. Inserting this expression
back into the equation for $\gamma$ we obtain a Ricatti equation, whose exact solution is \cite{Ince_56} 
\begin{equation}
 \gamma \left( t\right) = -2\sqrt{\rho_{0} \left(1-\rho_{0}\right)} e^{-\frac{t}{4}} \tanh{ 
\left[  \Xi \left ( \rho_0, t \right )   \right]} \label{4.12}
\end{equation}
with
\begin{equation}
\Xi \left ( \rho_0, t \right ) =  \tanh^{-1}{\left(\frac{\rho_{0}}{1-\rho_{0}}\right)^{1/2}} + \sqrt{\rho_{0} \left(1-\rho_{0}\right)} \left(  e^{-\frac{t}{4}} -1\right) .
\end{equation}
Since we have found explicit solutions for $\gamma$ and $\omega$ we can easily  revert to the original variables $\eta_+ = f  \gamma / \left ( \gamma + \omega \right)$
and $\eta_- = f  \gamma \left ( \omega - 1 \right) / \left ( \gamma + \omega \right)$ so as to write
\begin{equation}\label{eta+}
\eta_+ = - \rho_{0}\left(1-\rho_{0}\right)\left(2\rho_{0}-1\right) e^{-\frac{t}{4}} \sinh^{2}{\left[ \Xi \left ( \rho_0, t \right )  \right]}
\end{equation}
and
\begin{equation}
\eta_- =  - \eta_+ - \left [ \rho_{0} \left( 1 -\rho_{0} \right) \right]^{3/2} \left(2\rho_{0}-1\right)
 e^{-\frac{t}{2}} \sinh{\left[2 \Xi \left ( \rho_0, t \right )  \right]} .
\end{equation}
At this point we can immediately obtain $\alpha_+$, given in Eq.\ (\ref{4.06}), through a brute-force integration 
\begin{eqnarray}\label{4.15}
\hspace{-2.2cm} \alpha_{+} \left(t\right) &=& \frac{1}{2} \rho_{0} \left( 1 -\rho_{0} \right) \left(2\rho_{0}-1\right) e^{-\frac{t}{4}} \nonumber \\
\hspace{-2.5cm} && +\frac{1}{2} \rho_{0} \left( 1 -\rho_{0} \right) \left(2 - e^{-\frac{t}{4}}\right) \cosh{\left[ 2\sqrt{\rho_{0} \left( 1 -\rho_{0} \right)} \left(1 -e^{-\frac{t}{4}} \right)\right]} \nonumber \\
\hspace{-2.5cm} && - \frac{1}{2} \sqrt{\rho_{0} \left( 1 -\rho_{0} \right)} \left[1 - 2\rho_{0} \left( 1 -\rho_{0} \right) e^{-\frac{t}{4}} \right] \sinh{\left[2\sqrt{\rho_{0} \left( 1 -\rho_{0} \right)} \left(1 -e^{-\frac{t}{4}} \right) \right]}. 
\end{eqnarray}
Note  that $\alpha_{+} \left(t \to \infty \right) = \bar{y} \left ( \rho_0 \right)$ calculated within the 3-site approximation [see Eq.\ (\ref{3.18})].
Together with the expression for $f$ given in Eq.\ (\ref{4.08}), this result allows to obtain $\delta_+ = \alpha_+ - f$. Hence, to complete the solution of the system (\ref{4.06}) we need now
 to determine  $\alpha_-$ and $\delta_-$. This is achieved as follows.

The auxiliary variable $\chi = \frac{\alpha_{-}}{\alpha_{+}}$  satisfies 
\begin{equation}
\dot{\chi} = - \frac{1}{4} \left(1 + \chi\right) \frac{\eta_{+}}{\alpha_{+}}, \label{4.16}
\end{equation}
whose solution is simply
\begin{equation}\label{Xi}
\chi \left ( t \right ) = 2 \left ( 1 - \rho_0 \right ) e^{-G\left(t\right)} -1
\end{equation}
where we have used $\chi \left ( t = 0 \right ) = 1 - 2 \rho_0$ and
\begin{equation}\label{G}
G\left(t\right) = \frac{1}{4} \int_{0}^{t} \frac{\eta_{+}  \left(t^{\prime}\right) }{\alpha_{+}\left(t^{\prime}\right)} dt^{\prime}.
\end{equation}
Hence   $\alpha_-\left ( t \right ) = \alpha_+\left ( t \right )\chi \left ( t \right ) $ where the factors in the product are given by Eqs.\ (\ref{4.15}) and (\ref{Xi}).
To solve for the last unknown  $\delta_{-}$  we resort to a shortcut. Considering the definitions of $\delta_-$ and $\alpha_-$  in terms of the elementary probabilities
given in (\ref{4.01}) we see that they are related by the transformation $1 \leftrightarrow 0$ and so $\delta_{-} \left(t, \rho_{0}\right) = \alpha_{-} \left(t, 1-\rho_{0}\right)$.
This concludes the solution of the system (\ref{4.06}), but we note that  since we are not able to solve analytically the integral in Eq. (\ref{G}) the situation here is not as
satisfying as for the 3-site approximation. Most fortunately, that integral does not appear in the expressions for the expectations involving less than 4 sites, as we will see next.

\subsubsection{Calculation of $\rho$, $\phi$, $\psi$ and $w_1$.}
 Knowledge of these expectations will allow us to compare the $4$-site approximation predictions
with those of the $3$-site approximation. The simplest and most important of these expectations is $\rho$ which
can readily be written in terms of the previously introduced  variables 
\begin{equation}
\rho \left(t\right) = \alpha_{+}\left( t\right) + \delta_{+}\left( t\right) + \eta_{+}\left( t\right) + z_{1} \left( t\right) + w_{1} \left( t\right).  \label{4.18}
\end{equation}
To proceed further we need to derive explicit expression for  $w_{1} \left( t\right)$.
There is a simple and elegant way to do that other than replacing the r.h.s. of the equation for $\dot{w}_1$  with known quantities and then
integrating over $t$. In fact, introducing $\varepsilon \equiv z_{1} + w_{1}$ we have 
\numparts 
\begin{eqnarray}
\dot{\varepsilon} & =  & \frac{1}{4} \left(2 \eta_{+} -\eta_{-}\right)  \label{var1}  \\
 			       & =  & -\dot{\eta}_{+} - 2\dot{\alpha}_{+}, \label{var2}
\end{eqnarray}
\endnumparts 
where Eq.\ (\ref{var1}) was obtained by the direct substitutions $z_1 = \left ( \alpha_+ - \alpha_- \right )/2$, $z_2 = \left ( \eta_+ - \eta_- \right )/2$, $ \alpha_- = y_4 - z_1$
and $ \alpha_+= y_1 + z_2$ into the equations of $\dot{z}_1$ and $\dot{w}_1$ in system (\ref{4.02}). As for Eq.\ (\ref{var2}), it follows directly from the equation for
$\dot{\eta}_+$ and $\dot{\alpha}_+$ in system (\ref{4.06}). Direct integration of $\dot{\varepsilon}$ yields
\begin{equation}
\varepsilon \left ( t \right ) =  \varepsilon \left ( t= 0 \right )  - \eta_+ \left ( t = 0 \right ) - 2 \alpha_{+}\left ( t = 0 \right )  - \eta_+ \left ( t  \right ) - 2 \alpha_{+}\left ( t \right ) 
\end{equation}
which leads to
\numparts 
\begin{eqnarray}
\rho \left(t\right)  & = & \varepsilon \left ( t= 0 \right )  - \eta_+ \left ( t = 0 \right ) - 2 \alpha_{+}\left ( t = 0 \right ) -f \left( t\right)  \label{rho4_1} \\
			      & =  &\rho_{0}^{2} \left(3 - 2\rho_{0} \right) + \rho_{0} \left( 1 -\rho_{0}\right) \left( 1 - 2\rho_{0}\right) e^{-\frac{t}{4}} \label{rho4_2}
\end{eqnarray}
\endnumparts 
where we have used $f =  \alpha_+ - \delta_+$ given in Eq.\ (\ref{4.08}) and
\begin{equation}\label{eps0}
\varepsilon \left ( t= 0 \right )  - \eta_+ \left ( t = 0 \right ) - 2 \alpha_{+}\left ( t = 0 \right ) = \rho_{0}^{2} \left(3 - 2\rho_{0} \right)  = \bar{\rho} \left( \rho_{0} \right) .
\end{equation}
 Most remarkably, Eq.\ (\ref{rho4_2})  is identical to its counterpart for the $3$-site approximation
Eq. (\ref{3.05}) except for the argument of the exponential in which  $t/3$ is replaced by $t/4$.

To derive the 2-site expectation  $\phi = \left \langle \sigma_i \sigma_{i+1} \right \rangle $ we use the relation $ \phi = y_1 + z_2 + z_1 + w_1  = \alpha_+ + \varepsilon$
so that $\phi$  can be immediately derived  using the equations for $\alpha_+$ and $\eta_+$, Eqs.\ (\ref{4.06}) and (\ref{eta+}). In this case,  the form of the dependence 
of $\phi$ on $\rho_0$ and $t$ has no resemblance with the 3-site approximation counterpart, but the asymptotic result is exactly  the same. This can be seen by noting
that $\eta_+ \left ( t \to \infty \right ) = 0$ and so $\bar{\phi} = \bar{\rho} \left( \rho_{0} \right)  - \alpha_+  \left ( t \to \infty \right ) $ which is identical
to Eq.\ (\ref{phi_3}) since $\bar{y}\left( \rho_{0} \right)  = \alpha_+  \left ( t \to \infty \right )$.

The calculation of the 3-site expectation  $\psi = \left \langle \sigma_i \sigma_{i+1} \sigma_{i+2}  \right \rangle $ is equally simple. We use
the relation $\psi = z_{1} + w_{1} = \varepsilon$. Hence $\bar{\psi} = \bar{\rho}  \left( \rho_{0} \right)  - 2 \alpha_+  \left ( t \to \infty \right ) $ which
is identical to Eq.\  (\ref{psi_3}).

\begin{figure}[!h]
\centering\includegraphics[width=0.6\linewidth]{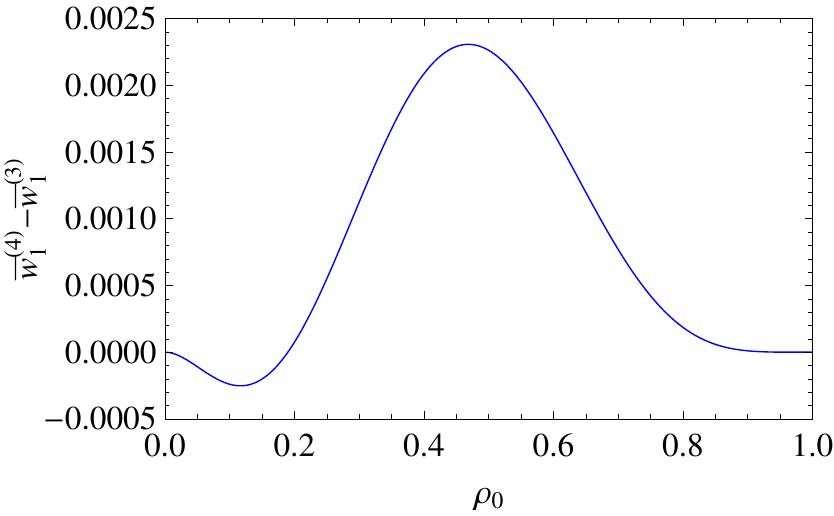}
\caption{Discrepancy between the steady-state expectation $\bar{w}_1 = \left \langle \sigma_i \sigma_{i+1} \sigma_{i+2} \sigma_{i+3} \right \rangle $ calculated with the 
$4$-site ($w^{(4)}_1$) and the $3$-site ($w^{(3)}_1$) approximations.}
\label{fig:4}
\end{figure}

The coincidence between the predictions of the $3$-site and $4$-site approximations for expectations involving 3 contiguous sites provides strong
evidence that those expectations are the exact results. However, this  agreement fails when considering expectations involving 4 or more contiguous sites as
we can appreciate by calculating $w_1 = \left \langle \sigma_i \sigma_{i+1} \sigma_{i+2}  \sigma_{i+3} \right \rangle$. We have
$w_1 = \varepsilon - z_1 = \varepsilon - \alpha_+ \left ( 1 - \chi \right )/2$.  Using Eq.\ (\ref{Xi}) for $\chi$  and taking $t \to \infty$ yields
\begin{equation}\label{4.26}
\bar{w}_{1} \left(\rho_{0}\right) = \bar{\rho} \left ( \rho_{0} \right )  +\bar{y} \left( \rho_{0} \right) \left[\left(1 -\rho_{0}\right)e^{-\bar{G} \left( \rho_{0} \right)} -3\right] 
\end{equation}
where $ \bar{G} \left( \rho_{0} \right) = \int_0^\infty  \left [ \eta_{+}  \left(t^{\prime}\right) / \alpha_{+}\left(t^{\prime}\right) \right ]  dt^{\prime}$
and  $\bar{y} \left( \rho_{0} \right) $ is given by Eq.\ (\ref{3.18}). As already pointed out we have to resort to a numerical integration to evaluate $\bar{G}$. 
Figure \ref{fig:3} shows the four expectations calculated in this section. In order to highlight the failure of the $3$-site approximation to
estimate 4-site expectations we present  in Fig.\ \ref{fig:4} the comparison between the 4-site  and the 3-site estimates of $\bar{w}_{1}$. The tiny difference
is imperceptible in the scale of Fig.\ \ref{fig:3} but it is sufficient to discard the possibility that $3$-site approximation is the exact solution of the majority-vote model.

\subsubsection{Probability of clusters of length $m$.}

The procedure here is identical to that applied for the $3$-site approximation, namely, decompose $P_{m+2} \left ( 1-\sigma, \sigma, \ldots, \sigma, 1- \sigma \right )$ with
$\sigma = 0,1$ in term of the elementary 4-site probabilities. The new element is that clusters of length $m=2$ can now be described directly by these elementary probabilities,
$P^{(4)}_{cl} \left (\rho_0,  2 \right )  = y_3 + y_4$ [see Eq.\ (\ref{4.01})] and yield 
\begin{equation}\label{P4cl2_2}
P^{(4)}_{cl} \left (\rho_0,  2 \right ) = \bar{y}\left (\rho_0 \right )   \left [ \rho_0 e^{-\bar{G}\left(1-\rho_{0}\right)} + \left ( 1 - \rho_0 \right )  e^{-\bar{G}\left(\rho_{0}\right)}
\right ] .
\end{equation}
The probability of clusters of length $m > 2$ is given by
\begin{equation}\label{P4cl2}
P^{(4)}_{cl} \left (\rho_0,  m \right ) = \frac{\left( \bar{\psi} - \bar{w}_{1}\right)^{2}}{\bar{\psi}} \left( \frac{\bar{w}_{1}}{\bar{\psi}} \right)^{m-3} + \frac{\left( \bar{\psi}_{-1} -\bar{w}_{-1}\right)^{2}}{\bar{\psi}_{-1}} \left( \frac{\bar{w}_{-1}}{\bar{\psi}_{-1}} \right)^{m-3}
\end{equation}
where $\bar{\psi}_{-1} \left ( \rho_0 \right ) = \bar{\psi} \left ( 1 - \rho_0 \right )$ and $\bar{w}_{-1} \left ( \rho_0 \right ) = \bar{w}_{1} \left ( 1 - \rho_0 \right )$ with
$\bar{\psi}$ and $\bar{w}_1$ given by Eqs.\  (\ref{psi_3}) and (\ref{4.26}), respectively. These probability distributions are exhibited in Figs.\ \ref{fig:5} and \ref{fig:6}. We find a perfect fitting
of the simulation data for $m=2$; for $m >2$ the fitting is good but there are  discrepancies in the vicinity of $\rho_0 = 0.5$, which are not perceptible in the scale
of the figures.

\subsubsection{Two-site correlations.} Since the results of the $4$-site approximation reduces to those of the $3$-site for expectations involving up to three contiguous
sites, correlations such as $\textrm{corr} \left(\sigma_{i},\sigma_{i+1}\right)$  and $\textrm{corr} \left(\sigma_{i},\sigma_{i+2}\right)$ are the same as for
the $3$-site approximation. In addition $\left \langle \sigma_{i} \sigma_{i+3} \right \rangle_{\left(4\right)} = w_1 + y_3$ and so
\begin{equation}\label{corr_3_44}
\left \langle \sigma_{i} \sigma_{i+3} \right \rangle_{\left(4\right)} = \bar{\rho} \left ( \rho_{0}\right) + \bar{y} \left(\rho_{0}\right) 
\left[\left(1 -\rho_{0}\right)e^{-\bar{G}\left(\rho_{0}\right)} +\rho_{0}e^{-\bar{G} \left(1 - \rho_{0}\right)} -3\right] .
\end{equation}
However, we need  the decomposition  in terms of the elementary 4-site probabilities to calculate expectations involving more  distant sites, such as 
\begin{equation}\label{corr_4_4}
\left \langle \sigma_{i} \sigma_{i+4} \right \rangle_{\left(4\right)} = \frac{\bar{w}_{1}^{2}}{\bar{\psi}} + \frac{\left(\bar{\psi}_{-1} - \bar{w}_{-1}\right)^{2}}{\bar{\psi}_{-1}} + 2\rho_{0} \bar{y}\left(\rho_{0}\right) e^{-\bar{G}\left(1 -\rho_{0}\right)} .
\end{equation}
These correlations are shown in Figs.\ \ref{fig:7}. Following the already observed pattern, we find a perfect agreement with the simulation data for quantities whose
calculation  involves expectations of  up to 4 contiguous sites.

\section{Discussion}\label{sec:Discussion}

Although the main purpose of this contribution is to show the remarkably good predictions of the $3$ and $4$-site approximations to describe the steady-state
properties of the extended one-dimensional majority-vote model, here we focus on the discussion of those properties rather than on the procedure to derive them.

\begin{figure}[!t]
  \begin{center}
\subfigure{\includegraphics[scale=0.75]{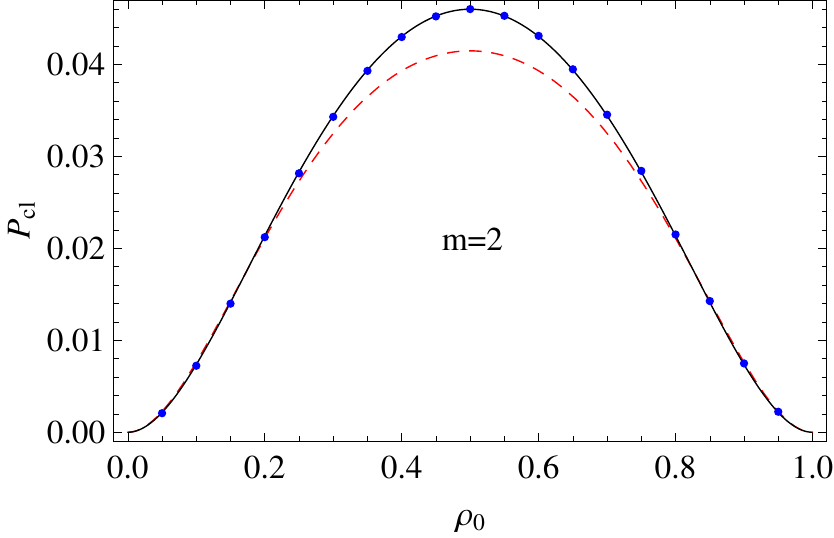}}
 \subfigure{\includegraphics[scale=0.75]{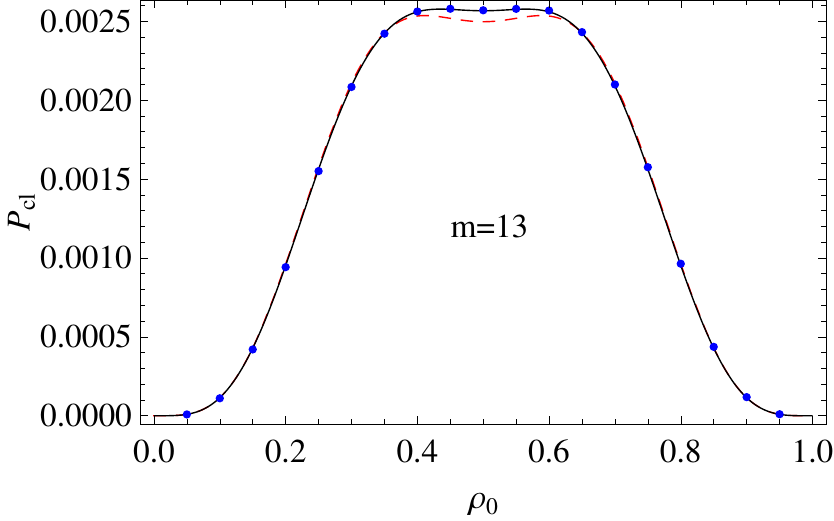}} \\
\subfigure{\includegraphics[scale=0.75]{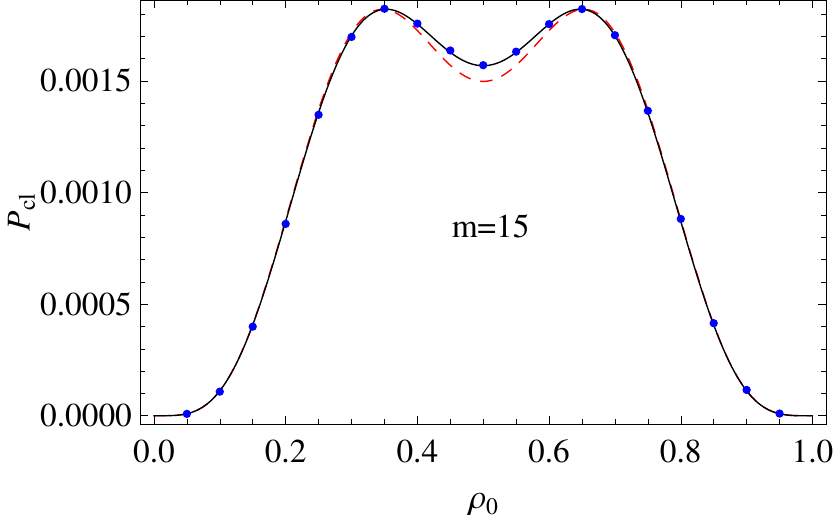}}
 \subfigure{\includegraphics[scale=0.75]{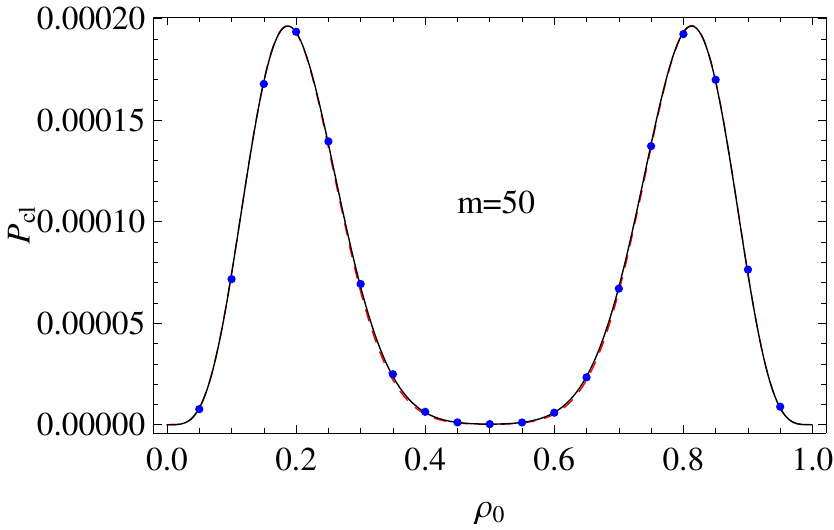}} 
  \end{center}
\caption{(Color online) Probability of finding clusters of length $m=2, 13, 15$ and $50$ as indicated in the figures.
The dashed curves are the results of the $3$-site approximation and the solid curves of the $4$-site approximation. The filled circles are the simulation data
for a ring of size $L=10^4$ and $10^6$ independent samples; the error bars
are smaller than the symbol sizes. The transition from a unimodal to a bimodal distribution takes place at $m= 13$.}
\label{fig:5}
\end{figure}

Figure \ref{fig:5} presents the probability of an absorbing configuration exhibiting  a cluster of length $m$  as function of the fraction of $1$s in the random initial
configuration. The $3$-site approximation does not provide a good quantitative account of the simulation data but it does provide an excellent qualitative picture
which captures the change of $P_{cl}$  from unimodal to bimodal that takes place for $m=13$. As for the $4$-site approximation, it provides a very good quantitative representation of the
simulation  data. In fact, the fitting  is perfect for  $m=2$ only, but the discrepancies are so small for $ m > 2$ that they are barely visible in the scale of the
figure.
We note that the transition of the distribution $P_{cl} \left (\rho_0,  m \right ) $ from unimodal  to bimodal was expected. In  fact, long clusters should be abundant  for
initial configurations with  $\rho_0$ close to $1$ or $0$ and very rare when the number of 1s and 0s is  well balanced as for $\rho_0 = 0.5$. What is surprising is
that the transition occurs for relatively large  $m$, indicating,  for example, that   it is more likely to find clusters of length 10 by starting with a balanced initial
configuration than with an unbalanced one.  Our simulations indicate that the transition from unimodal to bimodal takes place at $m = 13$ 
already for $L > 17$.  In addition, the distribution of cluster lengths is unimodal for $L \leq 9$ for all $2 \leq m \leq L-2$.

\begin{figure}[!t]
\begin{center}
\subfigure{\includegraphics[scale=0.75]{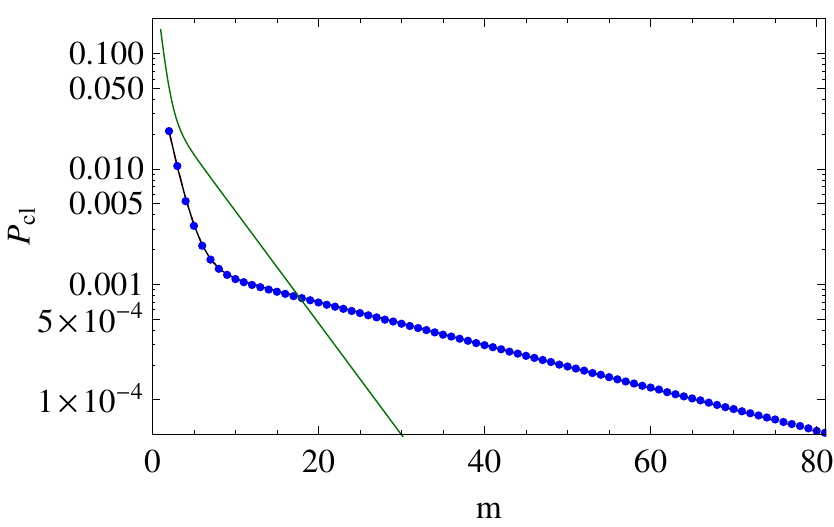}}
 \subfigure{\includegraphics[scale=0.75]{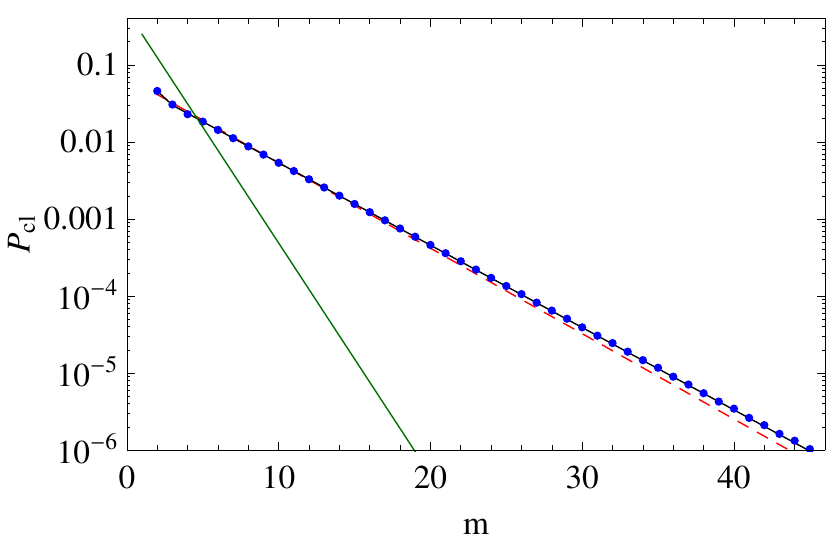}} 
  \end{center}
\caption{(Color online) Probability of finding clusters of length $m$ for fixed $\rho_0 = 0.2$ (left panel) and $\rho_0 = 0.5$ (right panel). 
The $3$-site approximation (dashed curves) and $4$-site approximations (solid lines connecting the symbols) give results  which are indistinguishable from
the simulation data (filled circles) in the scale of the figure for $\rho_0 = 0.2$; only for very large clusters one can see a noticeable  discrepancy
between the data and the results of the $3$-site approximation for $\rho_0 = 0.5$.
 For the purpose of comparison, the solid curves exhibit the results for  randomly assembled configurations. The simulations were carried out 
 for a ring of size $L=10^4$ and $10^6$ independent samples.}
\label{fig:6}
\end{figure}

Figure \ref{fig:6} shows the dependence of $P_{cl} \left (\rho_0,  m \right ) $ on $m$ for fixed $\rho_0$. It is not possible to distinguish from the 
results of the  $3$ and $4$-site approximations and the simulation data for $\rho_0 = 0.2$ but some observable discrepancies appear between the $3$-site approximation
and the data for large $m$ in the case $\rho_0 = 0.5$. The distribution is given by the sum of two exponentials [see Eqs.\  (\ref{P3cl2}) and (\ref{P4cl2})]
that account for the different possibilities  of  occurrence  associated to clusters composed of 1s and 0s for $\rho_0 \neq 0.5$.  For $\rho_0 = 0.5$ the arguments of the two exponentials 
become identical and so we have a single exponential decay.
Clearly, for  $\rho_0 =0.2$  clusters composed of 1s are dominant for small $m$ whereas clusters of
0s dominate in the large $m$ regime. The slopes of the exponentials are  complicated  functions of $\rho_0$, which can be well-approximated by Eq.\ 
(\ref{P4cl2})  derived within the $4$-site approximation scheme. For comparison, Fig.\ \ref{fig:6} shows $P_{cl}$ for randomly assembled configurations
which is given by
\begin{equation}
P^{random}_{cl} \left (\rho_0,  m \right ) = \left ( 1 - \rho_0 \right )^2  \rho_0^{m} + \rho_0^{2} \left ( 1 - \rho_0 \right )^{m} .  
\end{equation}
%

\begin{figure}[htp]
  \begin{center}
    \subfigure{\includegraphics[scale=0.75]{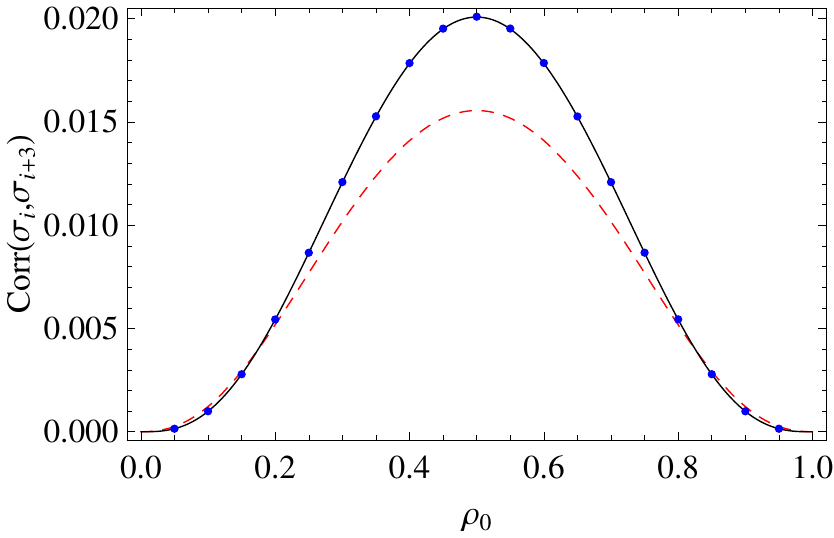}}
    \subfigure{\includegraphics[scale=0.75]{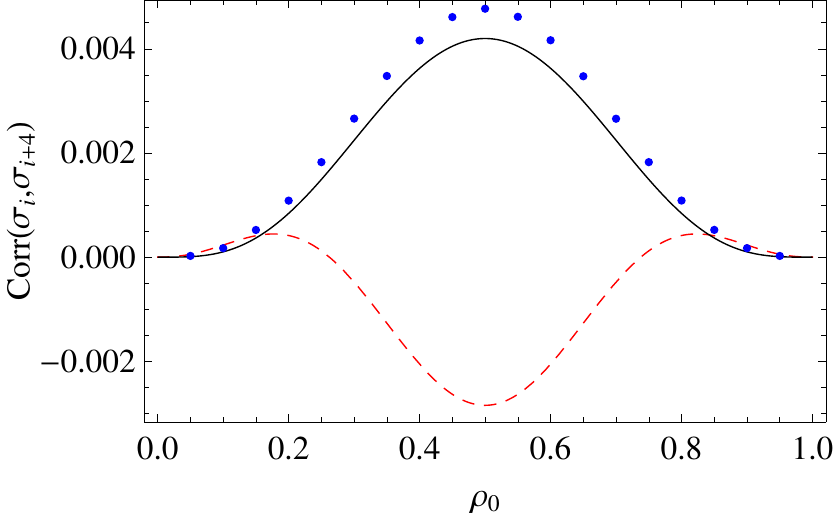}} 
  \end{center}
  \caption{(Color online) Comparison of the  results for the 2-site correlations obtained with the $3$-site approximation (dashed curves), the $4$-site approximation (solid curve)  and the
simulations (filled circles). The error bars are smaller than the symbol sizes. The left panel exhibits $\textrm{corr}\left(\sigma_{i},\sigma_{i+3}\right)$ for which the $4$-site approximation matches 
the data perfectly, and the right panel  shows $\textrm{corr}\left(\sigma_{i},\sigma_{i+4}\right)$ where we can see the failure of that approximation scheme. The simulations were carried out 
 for a ring of size $L=10^4$ and $10^6$ independent samples.}
  \label{fig:7}
\end{figure}

The failure of the $4$-site approximation in describing all the steady-state properties of the extended majority-vote model is better appreciated when
we consider the 2-site correlations, as shown in Fig.\ref{fig:7}. As already pointed out, the correlations $\textrm{corr}\left(\sigma_{i},\sigma_{i}\right)$,
$\textrm{corr}\left(\sigma_{i},\sigma_{i+1}\right)$ and $\textrm{corr}\left(\sigma_{i},\sigma_{i+2}\right)$ are described perfectly by both the
3 and 4-site approximations since they involve expectations of two and three contiguous  sites only, so Fig.\ \ref{fig:7} exhibits the more challenging
correlations,
$\textrm{corr}\left(\sigma_{i},\sigma_{i+3}\right)$ (left panel) and $\textrm{corr}\left(\sigma_{i},\sigma_{i+4}\right)$ (right panel). The $4$-site approximation
describes perfectly the former correlation but not the latter, whereas the $3$-site approximation fails in both cases. It is interesting that in all cases the 2-site correlations
exhibit a peak at $\rho_0 = 0.5$. This can be explained  by noting that the dynamics takes longer to freeze into one of the absorbing configurations for the well-balanced initial conditions
which results in highly correlated sites.  On the other hand, for $\rho_0$ close to its extreme values, most sites are already part of frozen  random clusters formed during the assemblage of
the initial configuration and so most of the sites at the final configuration  are uncorrelated.

Figure 7 shows, in addition, that the quality of the approximation  improves with increasing $n$, as expected. For example, estimation of $\textrm{corr}\left(\sigma_{i},\sigma_{i+4}\right)$
using the  3-site approximation (dashed curve) yields disastrous results, but the results obtained with the 4-site approximation (solid curve) are reasonable.  We note that
as $n$ increases  the estimation of statistical measures involving  $n+1$ sites  is improved. In fact, the relative error  resulting from use of the 3-site approximation to estimate
$\textrm{corr}\left(\sigma_{i},\sigma_{i+3}\right)$  at $\rho_0 = 0.5$ is 22.5\% whereas the error due to use of the 4-site approximation to estimate
$\textrm{corr}\left(\sigma_{i},\sigma_{i+4}\right)$ is 11.5\% only.

%

\section{Conclusion}

The extended one-dimensional majority-vote model is perhaps the simplest lattice model to exhibit an infinity of absorbing configurations. This
strong ergodicity breaking is probably the reason that  the  model is not  exactly solvable \cite{Privman_97,Marro_99}. From the mean-field theory 
perspective, which was the main  focus of our paper, the nontrivial nature of the steady state of the model  presented a most stimulating challenge
as the usual fixed-point equations proved quite uninformative. In fact, the solution to the problem is a one-to-one mapping between the randomly assembled initial configurations, which are described statistically by the 
density  $\rho_0$ of sites in state 1,   and the absorbing configurations. That mapping was obtained directly in the case of the pair approximation but
in the case of the $3$ and $4$-site approximations we had to solve analytically the dynamics for arbitrary $t$ and then take the asymptotic limit $t \to \infty$
in order to extract the mapping between the initial conditions and the steady state.

Although the pair approximation describes qualitatively the  mapping between $\rho_0$  and the statistical properties of the steady state, its predictions regarding
expectations involving two or more contiguous sites are not corroborated by the simulation results (see Fig.\ \ref{fig:2}).  
The 3-site approximation, however, produces a remarkable good
fitting of the simulation data for all quantities involving the expectation of  three contiguous sites. Moreover, the predictions of the $4$-site approximation reduce to those
of the $3$-site in the case of three contiguous sites expectations. We see this as a strong indication that the expectations
$\left \langle \sigma_i \right \rangle$,  $ \left \langle \sigma_i \sigma_{i+2} \right \rangle$ and $ \left \langle \sigma_i \sigma_{i+2} \sigma_{i+3}\right \rangle$ 
given by   Eqs.\ (\ref{3.06}), (\ref{phi_3}) and (\ref{psi_3}) are exact results. In addition, the perfect fitting of the simulation data by the expectation 
$\left \langle \sigma_i \sigma_{i+2} \sigma_{i+3} \sigma_{i+4}\right \rangle$, calculated within the $4$-site approximation [see Eq.\ (\ref{4.26})]
indicates that this quantity may be exact as well, but this evidence is not so strong as for the 3-site expectations.

The findings summarized above as well as a purely numerical analysis of the $5$ and $6$-site approximations (data not shown) reveal a most remarkable pattern:
the $n$-site approximation seems to yield the exact results for steady-state expectations involving
$n$ contiguous sites for $n>2$. We hope our paper will motivate further research to prove (or disprove) this assertion.

\ack The work of J.F.F. was supported in part by Conselho Nacional de
Desenvolvimento Cient{\'\i}fico e Tecnol{\'o}gico (CNPq)  and P.F.C.T. was supported by  
Funda\c{c}\~ao de Amparo \`a Pesquisa do Estado de S\~ao Paulo 
(FAPESP).

\end{document}